\documentclass[aps,prb,twocolumn,floatfix,superscriptaddress, citeautoscript]{revtex4}

\usepackage{graphicx}
\usepackage{amsmath}
\usepackage{amssymb}
\usepackage{color, soul}

\makeatother
\graphicspath{{./Images/}}

\newcommand{\beq}{\begin{equation}}
\newcommand{\eeq}{\end{equation}}
\newcommand{\beqy}{\begin{eqnarray}}
\newcommand{\eeqy}{\end{eqnarray}}
\newcommand{\beqs}{\begin{equation*}}
\newcommand{\eeqs}{\end{equation*}}
\newcommand{\bpm}{\begin{pmatrix}}
\newcommand{\epm}{\end{pmatrix}}
\definecolor{indigo}{rgb}{0.58,0.34,0.92}
\definecolor{agreen}{rgb}{0,0.5,0.2}

\newcommand{\vect}[1]{{\mathbf #1}}

\begin{document}

\title{Properties of the signal mode in the polariton OPO regime}

\author{K. Dunnett}
\email{kirsty.dunnett@su.se}
\affiliation{Nordita, KTH Royal Institute of Technology and Stockholm University, Roslagstullsbacken 23, SE-106 91 Stockholm, Sweden}

\author{A. Ferrier} 
\affiliation{ Department of Physics and Astronomy, University College London, Gower Street, London, WC1E 6BT, United Kingdom }

\author{A. Zamora} 
\affiliation{ Department of Physics and Astronomy, University College London, Gower Street, London, WC1E 6BT, United Kingdom }

\author{G. Dagvadorj} 
\affiliation{ Department of Physics and Astronomy, University College London, Gower Street, London, WC1E 6BT, United Kingdom }
\affiliation{ Department of Physics, University of Warwick, Coventry CV4 7AL, United Kingdom }

\author{M. H. Szyma\'nska}
\email{m.szymanska@ucl.ac.uk}
\affiliation{ Department of Physics and Astronomy, University College London, Gower Street, London, WC1E 6BT, United Kingdom } 

\date{\today}
 
\begin{abstract}
Theoretical analyses of the polariton optical parametric oscillator (OPO) regime often rely on a mean field approach based on the complex Gross-Pitaevskii equations in a three-mode approximation, where only three momentum states, the signal, pump and idler, are assumed to be significantly occupied. This approximation, however, lacks a constraint to uniquely determine the signal and idler momenta. In contrast, multimode numerical simulations and experiments show a unique momentum structure for the OPO states.  In this work we show that an estimate for the signal momentum chosen by the system can be found from a simple analysis of the pump-only configuration. We use this estimate to investigate how the chosen signal momentum depends on the properties of the drive.
\end{abstract}

\maketitle

\section{Introduction}

When exciton-polaritons in semiconductor microcavities are introduced by a coherent laser pump with energy $\omega_p$ and momentum $\mathbf{k}_p$ applied close to the inflection point of the lower polariton dispersion, pairs of pump polaritons can scatter to other states while conserving energy and momentum.  This is known as the optical parametric oscillator (OPO) regime.~\cite{PhysRevLett.85.3680, PhysRevB.63.041303, PhysRevB.63.193305, PSSB:PSSB200560961, SemicondSciTech.18.279}  Above a threshold pump strength, the parametric scattering leads to two new largely occupied states, the signal and idler, with energies $\omega_s, \omega_i$ and momenta $\mathbf{k}_s, \mathbf{k}_i$ respectively, satisfying $2\omega_p = \omega_s+ \omega_i$ and $2\mathbf{k}_p = \mathbf{k}_s + \mathbf{k}_i$.~\cite{PhysRevLett.85.3680, PhysRevB.63.041303, PhysRevB.63.193305} The large emission from the signal and idler modes can be of potential use in optical devices since the signal beam is both strong and directional.~\cite{PhysRevLett.85.3680, PhysRevB.62.16247}  Coherently pumped polariton systems have also been used to explore many-body collective phenomena such as polariton superfluidity\cite{amo2009superfluidity, sanvitto2010persistent, PhysRevB.92.035307} and non-equilibrium phase transitions.\cite{PhysRevX.5.041028}  

Theoretical analyses of the OPO regime commonly use a convenient three-mode description, where the dominant signal, pump and idler modes are the only ones present in the mean field.~\cite{PSSB:PSSB200560961, PhysRevB.71.115301, PhysRevB.75.075332, SemicondSciTech.18.279}  This description, however, suffers from an important deficiency: there are not enough constraints to uniquely determine the signal and idler momenta. Instead, for most system parameters, there is a range of momenta $\mathbf{k}_s$ and $\mathbf{k}_i$ at which the system is unstable to the OPO phase.\cite{PhysRevB.71.115301, Gavrilov2007} However, in experiments and numerical simulations of the full problem, there is no restriction on the number of modes that can be occupied and the system chooses a unique momentum structure which is usually dominated by a single $\mathbf{k}_s, \mathbf{k}_i$ pair.\cite{PhysRevB.65.081308, RevModPhys.85.299, Gavrilov2007, PhysRevLett.101.136401} 

Since the three-mode approximation is particularly convenient for investigating the OPO regime,~\cite{paper1, PhysRevB.75.075332, PhysRevA.76.043807} some means of determining which of the possible $\mathbf{k}_s$ values the system chooses would be useful.  While many previous works have explored the physics behind the formation and selection of the signal mode,\cite{PhysRevB.71.115301, EPL.67.997, Gavrilov2007, PhysRevLett.101.136401, PhysRevB.90.205303} ultimately, determining specifically which $\mathbf{k}_s$ is chosen analytically remains an unresolved problem.  Here we explore this open question by considering the instabilities of the below threshold description of the coherently pumped polariton system that includes only the pump mode, similar to methods employed in previous studies of the polariton OPO regime.\cite{PSSB:PSSB200560961, PhysRevB.71.115301, EPL.67.997, PhysRevA.76.043807, PhysRevB.74.245316, PhysRevB.75.075332, Gavrilov2007}  In particular, we find the momentum at which the pump mode is most unstable, i.e. which eigenstates have the largest positive imaginary part, and compare this with numerical solutions of the multimode problem, where the occupations show peaks at unique values of $\mathbf{k}_s$ and $\mathbf{k}_i$.  Throughout this paper we restrict our analysis to the optical limiter regime, where the pump-only solution is single valued, which is relevant to many recent studies pertaining to order and superfluidity in the polariton OPO regime,\cite{sanvitto2010persistent, PhysRevX.5.041028, PhysRevX.7.041006, PhysRevB.92.035307} as opposed to the bistable regime considered in many previous experimental and theoretical studies.\cite{EPL.67.997, PSSB:PSSB200560961, PhysRevB.75.075332, PhysRevB.71.115301, PhysRevA.69.023809, PhysRevB.90.205303, PhysRevB.68.115325, PhysRevB.77.115336, Gavrilov2007, PhysRevLett.101.136401}   

We first use a simplified lower polariton model to determine the polariton OPO phase diagram finding regions of pump intensity and possible $\mathbf{k}_s$ where (i) the pump-only  solution is unstable towards the OPO regime and (ii) the simple three-mode OPO description is stable to small fluctuations. The states which lie within region (i) but outside of region (ii), i.e. where both single and three-mode solutions are dynamically unstable, are characterised by more complicated momentum space structures not limited to three modes. 

In order to explore the characteristic patterns adopted by the polariton system, we solve numerically the equivalent multimode problem in various regimes across the OPO region. It turns out that, despite the fact that the OPO state can have a very complex momentum distribution, in almost all cases the largest (after the pump) occupations come from two unique modes, which we identify as the signal and the idler states with associated momenta $\mathbf{k}_s$ and $\mathbf{k}_i$. We then show that these momenta can be estimated by finding the most unstable mode in the simple single-mode approximation. Finally, we consider a two-component exciton-photon model (which gives a more precise description of the polariton system), and show that similar conclusions hold.  The exciton-photon model is then used to investigate how changing the pump momentum and energy (relative to the lower polariton dispersion) affects the resulting signal.

\section{Lower polariton model: one mode and three-mode descriptions \label{LowerPolaritons}}

In this section, we present the one- and three-mode approximations to the lower polariton field $\psi$ describing two-dimensional (2D) microcavity-polaritons, whose dynamics are given by
a momentum space complex Gross-Pitaevskii equation (cGPE) (with $\hbar=1$): 
 \begin{multline}
  i\partial_t \psi(\vect{k})^{} = \left[\omega_{lp}(\mathbf{k})
    -i\kappa_{lp} \right]\psi(\vect{k})^{}  + \\ \sum_{\vect{k}_1, \vect{k}_2} g_{\vect{k},
    \vect{k}_1, \vect{k}_2} \psi^*(\vect{k}_1 + \vect{k}_2-\vect{k})
  \psi(\vect{k}_1)^{} \psi(\vect{k}_2)^{}  + F_pe^{-i\omega_p t}
  \delta_{\vect{k},\vect{k}_p}\; .
\label{LPcGPE}
\end{multline} 
The lower polariton dispersion is $\omega_{lp}(\mathbf{k}) = \frac{1}{2}\left( \omega_c(\mathbf{k})+\omega_x -\sqrt{(\omega_c(\mathbf{k})-\omega_x)^2+\Omega_R^2} \right)$. $\omega_x$  and $\omega_c(\mathbf{k})$ are the exciton and cavity photon dispersion relations respectively, $\Omega_R$ is the Rabi frequency of the exciton-photon coupling, $\kappa_{lp}$ is the polariton decay rate and $g_{\vect{k}, \vect{k}_1, \vect{k}_2}=g_X X_{\vect{k}} X_{\vect{k}_1 + \vect{k}_2-\vect{k}} X_{\vect{k}_1} X_{\vect{k}_2}$ describes the momentum dependent polariton-polariton interactions with Hopfield coefficients $X_{\mathbf{k}}$,~\cite{PhysRevB.75.075332, PhysRevA.76.043807, RevModPhys.85.299} and exciton-exciton interaction $g_X$. To facilitate the numerical integration of the polariton cGPE, we consider a simplified situation where the polariton-polariton interaction is treated as being constant. The mean field and linear response results are close to those obtained when the full momentum dependence is included and are therefore representative.~\cite{KDunnettThesis}  The external pump $F_{p}$ introduces polaritons at the pump energy $\omega_p$ and momentum $\mathbf{k}_p$.  

In the following we use dimensionless units, measuring time, length, and energy in units of $2/\Omega_R$, $\sqrt{\hbar/(\Omega_Rm_c)}$ and $\hbar\Omega_R/2$ respectively, where $m_c$ is the cavity photon effective mass.  Without any loss of generality we can choose the pump wave-vector to be along the $x$ direction i.e. $\mathbf{k}_p=(k_p,0)$.  Unless stated otherwise, we use $\mathbf{k}_p = (1.4,0)$, resonant to the lower polariton dispersion with $\omega_p = \omega_{lp}(k_p) = -0.42$, in dimensionless units where $\omega_c(0) = \omega_x = 0$, and use a polariton decay rate of $\kappa_{lp} = 0.045$.

\subsection{Single mode description and estimate of the signal momentum \label{sec:one_mode_pola}}

The first step in the theoretical analysis of the OPO transition is to
determine the mean field solution assuming that only one mode, 
coinciding with the pump energy and momentum, is occupied.~\cite{PhysRevB.75.075332, PhysRevB.71.115301, PSSB:PSSB200560961} This means using the following ansatz 
\beq
\psi(\vect{k})^{} = P e^{-i\omega_p t}\delta_{\mathbf{k},\mathbf{k}_p}. \label{PumpMF}
\eeq 
In the lower polariton model, the mean field occupation of the pump mode and the eigenvalues of the linear response matrix can be calculated exactly.~\cite{PhysRevB.75.075332, PSSB:PSSB200560961, PhysRevB.71.115301}  Substituting Eq. \eqref{PumpMF} into the cGPE Eq. \eqref{LPcGPE} and considering the mean field steady state
$i\partial_t P = 0$ gives 
\beqs 
F_p = (\omega_{lp}(\mathbf{k}_p) -\omega_p +g_p |P|^2  -i\kappa_{lp})P, 
\eeqs 
where $g_p \equiv g_X X_{\mathbf{k}_p}^4$. Thus, the mean field occupation in the pump mode ($n_p = |P|^2$) becomes 
\beqs
|F_p|^2 = ((\omega_{lp}(\mathbf{k}_p) -\omega_p + g_p n_p)^2 +\kappa_{lp}^2)n_p.
\eeqs 
The mean field already gives some information about the expected behaviour: $|F_p|^2$
is cubic in $n_p$ which can lead to bistable behaviour under certain pumping parameters. 
The critical quantity is the detuning of the pump away from the lower polariton dispersion: 
if $\omega_p - \omega_{lp}(\mathbf{k}_p) > \sqrt{3} \kappa_{lp}$, the pump mode is bistable.\cite{PSSB:PSSB200560961, PhysRevB.75.075332, PhysRevB.71.115301, PhysRevA.69.023809, PhysRevB.90.205303} However, in this work we do not consider the bistable regime but restrict ourselves to the optical limiter regime, with a monotonic relation between the pump strength and the polariton occupation $n_p$.  

The next step is to perform a linear response analysis (linear Bogoluibov-like theory ~\cite{PSSB:PSSB200560961, RevModPhys.85.299}) by expanding in fluctuations around the pump mode, which means using a new ansatz:~\cite{PhysRevB.75.075332, PhysRevB.71.115301, PSSB:PSSB200560961, RevModPhys.85.299} 
\beqs
\psi = P e^{-i\omega_p t}\delta_{\mathbf{k},\mathbf{k}_p} + \Delta
P e^{-i(\omega_p+\omega)t}\delta_{\mathbf{k},\mathbf{k}_p+\Delta \mathbf{k}}, 
\eeqs 
where $\omega$ is the energy and $\Delta \mathbf{k}$ the momentum of the fluctuations.  The linear response matrix $L$ is formed by keeping only terms that are linear in fluctuations to give:~\cite{RevModPhys.85.299, PhysRevB.75.075332, PhysRevB.63.041303, SemicondSciTech.18.279}  
\beq
L = \bpm \alpha^+ -i\kappa_{lp} & g_p P^2 \\
-{g_p P^*}^2  & -\alpha^- -i\kappa_{lp}  \epm \label{LPLRmatrix},
\eeq 
where 
\beq
\alpha^\pm = \omega_{lp}(\mathbf{k}_p \pm \Delta \mathbf{k}) -\omega_p
+ 2g_p|P|^2. \label{alphadef} 
\eeq 
The matrix $L$ satisfies:~\cite{PhysRevB.75.075332, PSSB:PSSB200560961} 
\beqs
L(\mathbf{k}) \Delta\Psi = \omega(\mathbf{k}) \Delta\Psi, 
\eeqs 
where $\Delta\Psi$ is a vector of the fluctuations and $\omega(\mathbf{k})$ are the complex eigenvalues, the real parts of which give the spectra of the excitations, $\Re(\omega)$, while the imaginary parts give the regions of instability, \mbox{$\Im(\omega)>0$}.\cite{PSSB:PSSB200560961, PhysRevB.75.075332, PhysRevB.71.115301, EPL.67.997, PhysRevA.76.043807, PhysRevB.74.245316, Gavrilov2007} The eigenvalues of Eq. \eqref{LPLRmatrix}, are:~\cite{paper1, PhysRevB.75.075332}   
\beq
\omega^{\pm} = \frac{\alpha^+-\alpha^-}{2} -i\kappa_{lp} \pm
\frac{1}{2}\sqrt{(\alpha^++\alpha^-)^2 -4n_p^2}, \label{LPeivals} 
\eeq 
with $\alpha^\pm$ defined in Eq. \eqref{alphadef}. If the discriminant in Eq. \eqref{LPeivals} is positive, then the two eigenvalues have a common imaginary part, $\Im(\omega^\pm) = -\kappa_{lp}$, and the pump-only state is stable.   When the discriminant is negative, the imaginary parts of the eigenvalues differ and it is possible to find the location of the maximum of $\Im(\omega^+)$. This can become positive leading to instability towards some new solution, 
the simplest of which is the three-mode OPO state.~\cite{PhysRevB.75.075332, PSSB:PSSB200560961}  

\begin{figure}[t!]
\includegraphics[width=\columnwidth]{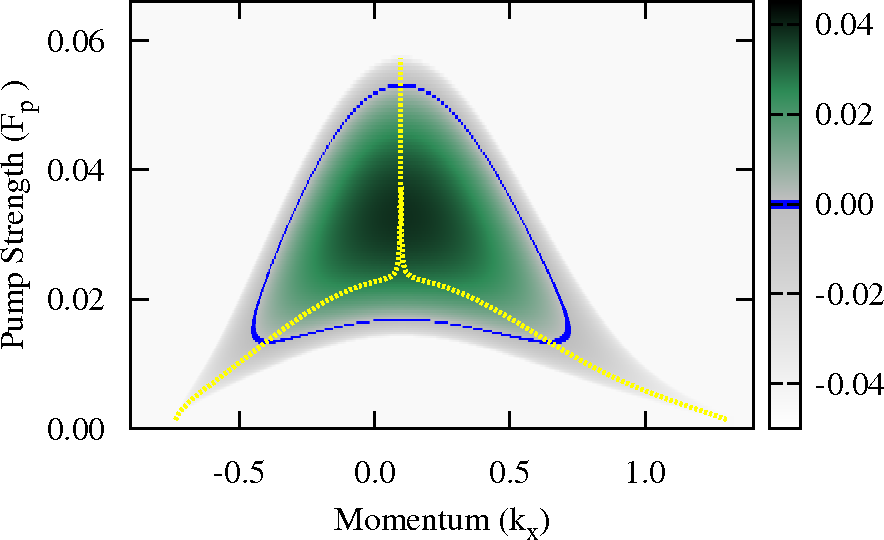}
\caption{ The imaginary part of the eigenvalues [$\Im(\omega^+)$] that become positive. Maximum values of $\Im(\omega^+)$ at each pump strength are marked by yellow dashed lines. The blue line ($\Im(\omega^+)$ = 0) encloses the region of instability. \label{fig:ksoverIms_X1}} 
\end{figure}

In Fig. \ref{fig:ksoverIms_X1} we plot $\Im(\omega^+)$ [Eq. \eqref{LPeivals}] as a function of the pump strength $F_p$ and momentum along the $x$ direction of the considered excitation. 
The blue line in Fig \ref{fig:ksoverIms_X1} indicates the contour on which $\Im(\omega^+)$ changes sign; i.e. the region inside is where this pump-only  solution is unstable. The momentum at which
$\Im(\omega^+)$ has a maximum value (which we call $k_s^m$) for a given pump strength is marked by the yellow dotted line and overlaid to highlight where the maxima lie in relation to the borders of the unstable region.

\begin{figure}[ht!]
\includegraphics[width=\columnwidth, height = 6.5cm]{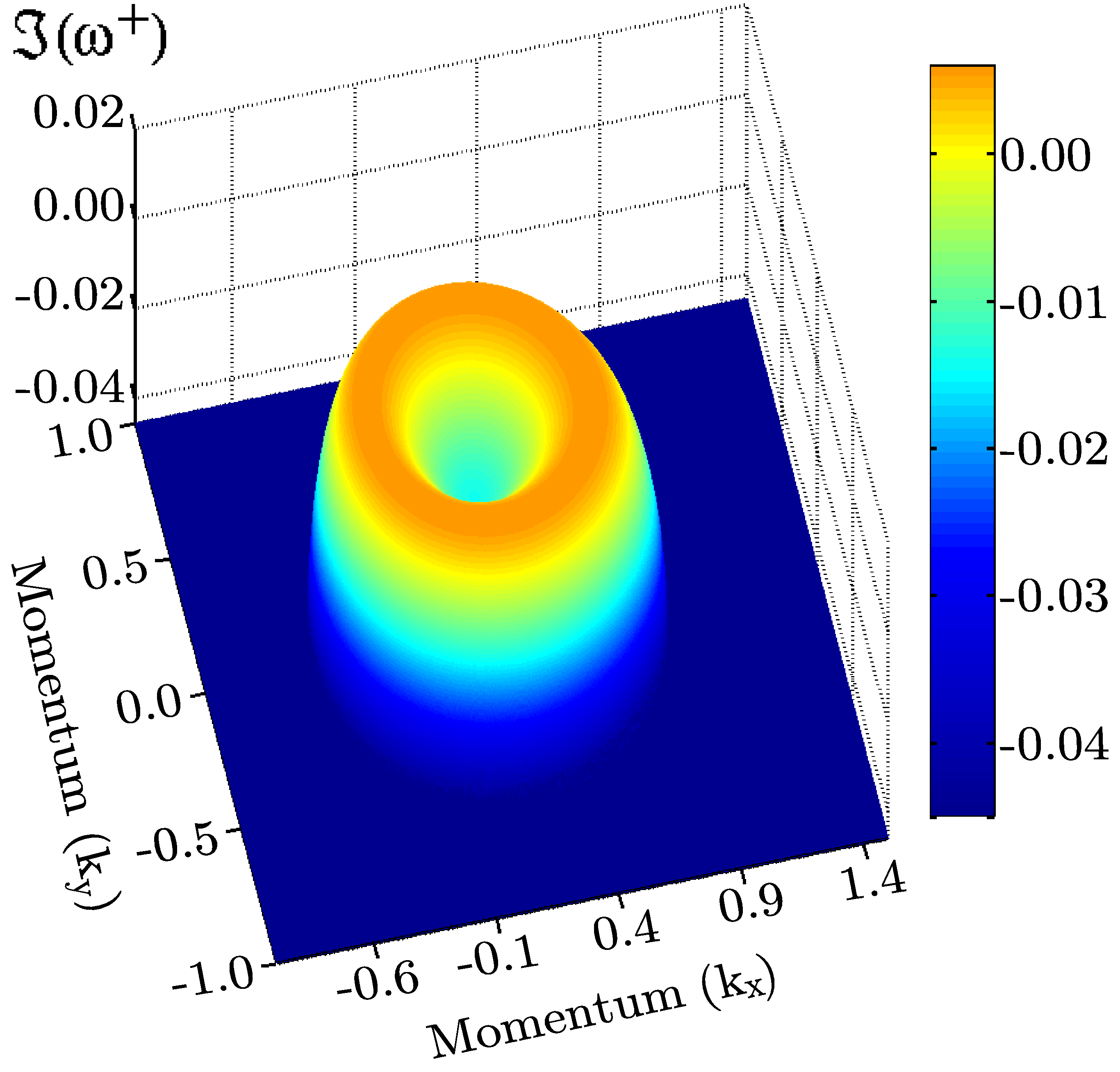}
\caption{ Imaginary part of the eigenvalues [$\Im(\omega^+)$] in 2-D momentum space $(k_x, k_y)$ for the pump-only state close to lower threshold ($F_p=0.0157$).  
The condition $\Im(\omega^+)>0$ determines a ring shaped region in k-space where the system can develop signal modes. \label{fig:circle} 
}
\end{figure}

At very weak pumping, the $k_s^m$ values form a ring in momentum space, an example of which is shown in Fig. \ref{fig:circle}.  This appears as two branches in Fig. \ref{fig:ksoverIms_X1} that indicate the $k_x$ values at which the ring at each $F_p$ crosses $k_y = 0$.  As the pump strength is increased, the maximum value of $\Im(\omega^+)$ increases and then decreases to again become negative. There is an intermediate pump strength at which the ring of $k_s^m$ values become indistinguishable; the $k_s^m$ values approach this `coalescence' point evenly. This coalescence point occurs at comparatively weak pumping, in the centre of the unstable region, but before the imaginary part reaches its maximum value. Once there is a single $k_s^m$, its value is constant until the discriminant of Eq. \eqref{LPeivals} becomes positive, after the pump mode has become stable again.

\subsection{Three-mode approximation}

To describe the OPO regime, which may occur inside the blue contour in Fig \ref{fig:ksoverIms_X1}, the simplest ansatz consists of three modes $\psi = \psi_s + \psi_p + \psi_i$, the signal, pump and idler respectively, with each mode being a plane wave with amplitude $M$:
\beqs
\psi_{m} = Me^{-i\omega_m t}\delta_{\mathbf k, \mathbf q_m}.  
\eeqs 
We substitute this into the lower polariton cGPE \mbox{[Eq. \eqref{LPcGPE}]}.\cite{PhysRevB.71.115301, PhysRevA.76.043807, PhysRevB.75.075332} Some of the interaction terms introduce modes outside of the three-mode ansatz; these are discarded by requiring energy and momentum conservation within the OPO modes. This leads to three coupled \mbox{cGPEs}, one for each of the signal, pump and idler modes. Considering the steady-state with \mbox{$\partial_t P = \partial_t S = \partial_t I = 0$} gives three complex equations~\cite{PhysRevB.71.115301, PhysRevA.76.043807, paper1} 
\beqy
{\Xi_s S + \tilde{g}_xP^2I^*} &=&0 , \label{1} \\
{\Xi_p P + 2\tilde{g}_xSP^*I +f_p} &=&0, \label{2}\\
{\Xi_i I + \tilde{g}_xS^*P^2} &=&0,  \label{3} 
\eeqy  
where $\tilde{g}_x \equiv g_X X^2_pX_iX_s$, with the shorthand notation $\Xi_j\equiv \omega_{\mathrm{lp}} (\mathbf{k}_j)-\omega_j-i\kappa_{lp}+g_X X^2_j\left(2(\tilde{n}_s+ \tilde{n}_p + \tilde{n}_i) -\tilde{n}_j\right)$ (with $\tilde{n}_j=X^2_jn_j=X^2_j |J|^2$) and the external pump $f_p = F_pe^{-i\omega_p t}\delta_{\mathbf{k},\mathbf{k}_p}$, with the redefinition $X_p F_p \to F_p$. Note that the above set of equations is invariant under a global U(1) phase rotation of the form: 
 \begin{align}
 \begin{pmatrix}
   S \\
   I \\
   P 
 \end{pmatrix}
\to
 \begin{pmatrix}
   S' \\
   I' \\
   P'
 \end{pmatrix}
=
 \begin{pmatrix}
  e^{i\alpha}    &    0    &   0 \\
    0            & e^{-i\alpha} &  0 \\
    0            &  0        &  1
 \end{pmatrix}
 \begin{pmatrix}
   S \\
   I \\
   P
 \end{pmatrix}; \nonumber
\end{align} 
 \begin{equation}
S\to S'=Se^{i\alpha}, \; I \to
I'=Ie^{-i\alpha} , \;\;\; 
P\to P'=P. \nonumber
\label{eq_sym_global}
\end{equation} 
The U(1) symmetry can be broken spontaneously, and is responsible for the existence of a gapless phase mode above the OPO threshold, leading to interesting universal critical phenomena such as long-distance coherence~\cite{PhysRevB.74.245316} and superfluidity.~\cite{PhysRevB.92.035307}  The U(1) symmetry allows us to freely choose the phase of one of the fields.  Specifically, we choose the signal $S$ to be real.  We are, however, still left with seven unknowns: the three amplitudes $|S|$, $|P|$, $|I|$; the two remaining phases; and the energy $\omega_s$ and momentum $\mathbf{k}_s$ of the signal mode (the idler mode is linked to the signal via energy and momentum conservation $2\omega_p = \omega_s+ \omega_i$ and $2\mathbf{k}_p = \mathbf{k}_s+\mathbf{k}_i$). With six real constraints given by Eqs. (\ref{1})-(\ref{3}) there remains one unknown not set by the three-mode mean field conditions. This is often chosen to be the signal momentum, $\mathbf{k}_s$. The main purpose of this work is to provide a method by which we can estimate the $\mathbf{k}_s$ that would be chosen by a real physical system, for example the one which occurs in the numerical solution of the full multimode problem.

The three-mode description of the polariton OPO regime gives a finite signal mode occupation $n_s$ for a wide range of $k_s$ and $F_p$, corresponding to the momenta and pump strengths at which the single mode description is unstable.  Therefore, for each signal momentum, $k_s$, and pump strength, $F_p$, we ask two questions: i) whether the OPO regime exists (non-zero $n_s$) and ii) whether it is stable to fluctuations in $k_x$ and $k_y$ (all imaginary parts of the eigenvalues $\leq 0$), since the existence of a mean field solution does not imply its stability.~\cite{paper1, PhysRevB.71.115301, PhysRevB.75.075332} 

The stability of the OPO state for a given pump strength and $\mathbf{k}_s$ is considered through linear response analysis:~\cite{PhysRevA.76.043807, paper1, PhysRevB.71.115301}  
\beqs
L_{OPO} = \bpm -M(\Delta \mathbf{k}) & -Q(\Delta \mathbf{k}) \\
Q^*(-\Delta \mathbf{k}) & M^*(-\Delta \mathbf{k}) \epm 
\eeqs
with the sub-matrices ~\cite{paper1, PhysRevA.76.043807}
\beqy 
M_{m,n}(\Delta \mathbf{k}) &=& \delta_{m,n}\left(\omega_m-\omega_{lp}(\mathbf{k}_m+\Delta \mathbf{k}) +i\kappa_{lp}\right)
\nonumber \\ 
&& - 2 \sum_{r,t=1}^3\delta_{m+r,n+t}g \psi_r^*\psi_t,\nonumber \\ 
Q_{m,n}(\Delta \mathbf{k}) &=& -\sum_{r,t=1}^3
\delta_{m+n,r+t} g \psi_r\psi_t, \nonumber 
\eeqy   
where $n,m,r,t=1,2,3=s,p,i$.  There are six eigenvalues, three of which have imaginary parts that are always less than $-\kappa_{lp}$, and three of which have imaginary parts that may become positive.  

\begin{figure}[h!]
\includegraphics[width=\columnwidth]{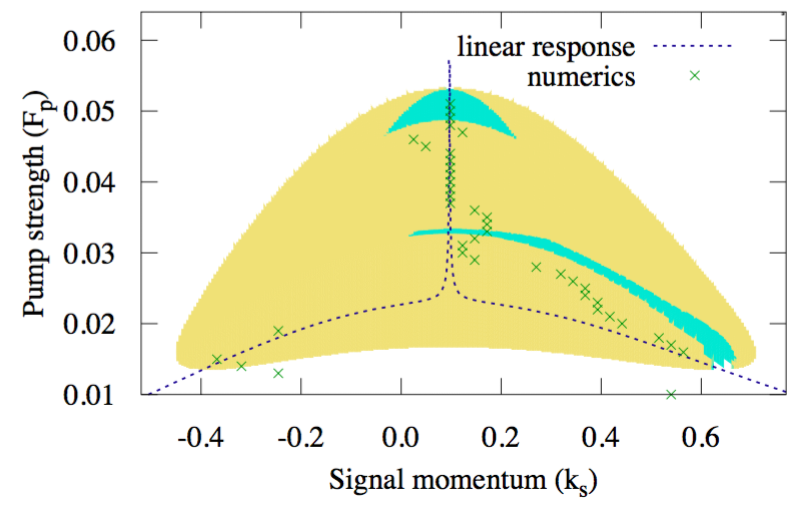} 
\caption{ Region where the single mode ansatz is unstable, with the areas where the three-mode ansatz is stable (turquoise) and unstable (yellow) to fluctuations in both $k_x$ and $k_y$. The blue dotted line indicates the momentum $k_s^m$ for which the imaginary part of the eigenvalues around the pump-only mean field has a maximum at a given $F_p$, and the green crosses mark the location of the signal found in the numerical simulations.  
\label{fig:OPO_kskykx}} 
\end{figure}

The phase diagram in Fig. \ref{fig:OPO_kskykx} shows the three types of behaviour that occur: stable single mode solution and no possibility for OPO (white region); both single and three-mode solutions being unstable (yellow region); stable three-mode OPO ansatz (turquoise regions).~\cite{PhysRevX.7.041006} We observe that at low pump strengths, three-mode OPO is stable for a range of $k_s$ on the side towards the pump. As the pumping is increased, this region narrows slightly and moves towards $k_s = 0$. Further increase of the pumping leads to a region where three-mode OPO is unstable for all $k_s$ for which it exists. At the highest pump strengths, there is again a region where three-mode OPO is stable.  This is centred on a small positive $k_s$ and includes $k_s = 0$. Only at the highest pump strengths does the $k_s^m$ from the single mode linear response analysis lie consistently within the region of stable three-mode OPO. The green crosses in Fig \ref{fig:OPO_kskykx} are the $k_x$ values of the most occupied modes from the numerical integration of the cGPE and are discussed fully in section \ref{sec:num}.

\begin{figure}[b]
\includegraphics[width=\columnwidth]{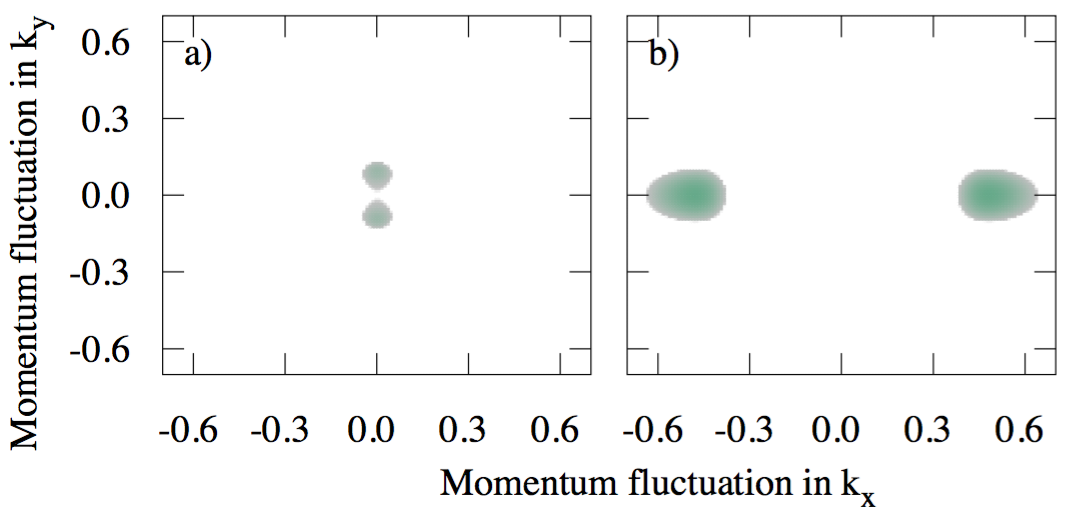}
\caption{ Unstable eigenvalues (positive imaginary part) around the OPO for fluctuations in $k_x$ and $k_y$.  $\mathbf{k}_s = (0,0)$: a) $F_p = 0.0316, n_p = 0.048$; b) $F_p = 0.0325, n_p = 0.052$.  
\label{fig:OPOkxkyinstabs}} 
\end{figure}

In Fig. \ref{fig:OPOkxkyinstabs} we plot two examples of the largest imaginary part of the eigenvalues of the linear response for specific unstable configurations of the three-mode ansatz, with the signal momentum $\mathbf{k}_s = \vect{0}$.  These two examples highlight that the three-mode description may be stable to fluctuations along the line $k_y = 0$ but show instabilities at finite $k_y\neq 0$. We observe different structures in the momentum fluctuation along the full phase diagram of the three-mode ansatz, some of which are quite complex.\cite{KDunnettThesis}

\section{Numerical solution of the multimode polariton field \label{sec:num}}

So far, the results presented have been obtained by considering a single mode or a three-mode ansatz for the polariton field. However, as highlighted by the large yellow region in Fig. \ref{fig:OPO_kskykx}, a complete description of the system requires the consideration of solutions beyond the three-mode ansatz, since the three-mode description of the OPO regime only gives stable solutions for a small part of the full OPO region. In this section we consider the multimode scenario by numerically integrating the non-linear dynamical equation of the lower polariton field given by Eq. \eqref{LPcGPE} in  2D real-space $\mathbf{r}=(x,y)$ and with a momentum-independent polariton-polariton interaction. Further, we describe the polariton field through a stochastic complex Gross-Pitaevskii equation (scGPE), which can be obtained from the mapping of the Fokker-Planck equation of the time-evolution of the quasi-probability function onto a Langevin equation on a grid with lattice spacing $a$; provided $g_X/(\kappa_{lp}dV)\ll 1$,\cite{PhysRevB.72.125335} where $dV = a^2$, the volume element of the considered lattice, controls the truncation condition of the Wigner representation. 
 
The scGPE for the lower polariton field in real-space reads:
\begin{align}
&i d \psi(\mathbf{r}) = \left\{ \hat{H}_{MF} \psi(\mathbf{r}) + f_p \right\}dt  + i\sqrt{\kappa_{lp}}dW,
\label{eq:sCGPE_pola}
\end{align}
where the complex valued, zero-mean, white Wiener noise $dW$ fulfils 
$\langle dW^{*} (\vect{r},t) dW (\vect{r}',t) \rangle =\frac{1}{dV}\delta_{\vect{r},\vect{r}'} dt$, the external drive $f_p = F_p e^{i (\vect{k}_p \cdot \vect{r} - \omega_p t)}$ and the operator $\hat{H}_{MF}$ abbreviates
\beqy
\hat{H}_{MF}= \omega_{lp}(-i\nabla)-i\kappa_{lp} + g_X|\psi(\mathbf{r})|^2_{-}.
\label{eq:H_pola}
\eeqy
We consider a constant interaction $g_X$ and $|\psi(\mathbf{r})|^2_{-} = |\psi(\mathbf{r})|^2  - \frac{1}{dV}$.~\cite{PhysRevB.72.125335}

\subsection{Mean field polariton distributions}

\begin{figure}[t]
\includegraphics[width=\columnwidth]{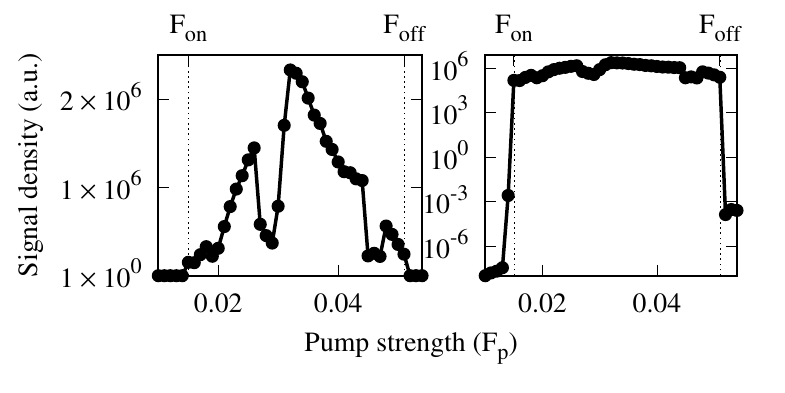}
\caption{Signal density in linear (left) and logarithmic (right) scales.  The lower and upper thresholds are identified as $F_{\mathrm{on}}$ and $F_{\mathrm{off}}$ respectively. \label{fig:X1_xmds_ks}} 
\end{figure}

First, we consider the mean field solution of the multimode and non-linear problem, by integrating Eq.\eqref{eq:sCGPE_pola} in time with the Wiener noise term $dW$ set to zero and replacing $|\psi(\mathbf{r})|^2_{-}$ by $|\psi(\mathbf{r})|^2$.  In both this model and later in the exciton-photon model, numerical integration is performed using the XMDS2 package,\cite{XMDS_CPhysComm.184.201} with either an adaptive step size Runga-Kutta algorithm (for mean field numerics), or a fixed step size semi-implicit algorithm with 5 iterations (for stochastic numerics), on a $256\times256$ lattice for a plane-wave external pump.  In all the numerical results reported here, we consider a Rabi splitting $\Omega_R = 4.4\mathrm{meV}$ and a cavity photon mass $m_c = 2.3\times 10^{-5} m_e$, as appropriate for typical experimental systems.\cite{sanvitto2010persistent}  We identify the OPO regime by finding a significant occupation of the signal mode, i.e. $|\psi_s|^2 > 10^4$, where the $\psi_s$ is the signal field at some $k_x < k_p$.  In Fig. \ref{fig:X1_xmds_ks} we show the signal density as a function of the pump strength, on both linear and logarithmic scales, where it can be seen that OPO appears for a range of pump strengths $F_{\mathrm{on}} \leq F_{p}\leq F_{\mathrm{off}}$.  In our units, we have $F_{\mathrm{on}}\approx 0.0135$ and $F_{\mathrm{off}}\approx 0.053$, which match closely to the OPO thresholds given by the onset of instability of the pump-only solution.  

\begin{figure}[t]
\includegraphics[width=\columnwidth]{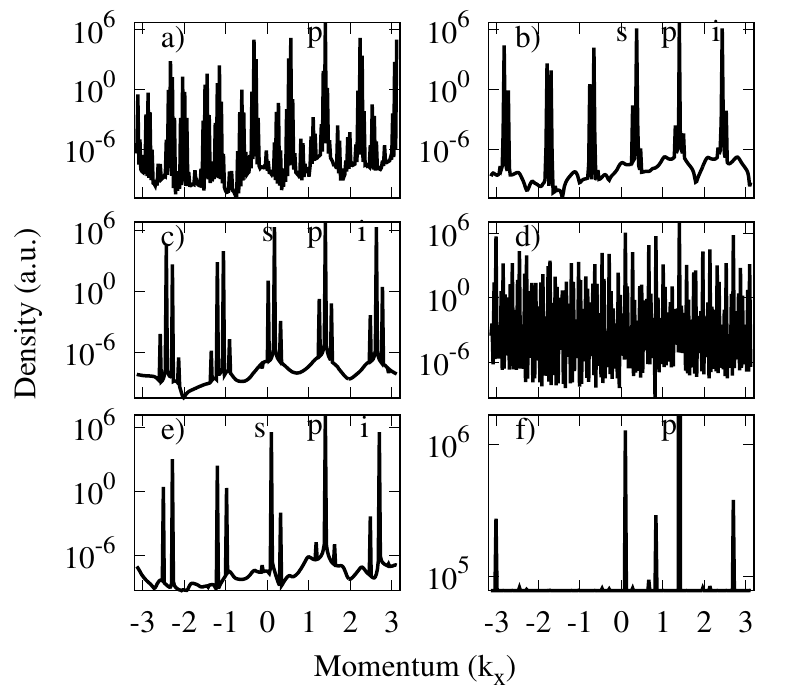}
\caption{Polariton density profiles at various pump strengths across OPO regime. a)-e) Log scale profiles: a) $F_p = 0.016$, b) $F_p = 0.025$, c) $F_p= 0.034$, d) $F_p = 0.043$, e) $F_p = 0.050$; f) $F_p = 0.043$, linear scale.  
The three-mode description of the OPO regime is unstable for $F_p= 0.043$ [d) and f)]. The labels p, s and i label the pump, signal and idler modes respectively, with the latter two only well defined for some $F_p$.
\label{fig:Densities_stable}} 
\end{figure}

The system in the steady-state exhibits two distinct behaviours.  First, we observe that for certain values of $F_p$, the full multimode polariton field in the steady-state conforms to the three-mode description, since we see the appearance of three significantly occupied modes at three distinct values of $k_x$ (signal $k_x<k_p$, pump $k_x=k_p$, and idler $k_x>k_p$, respectively), with occupations orders of magnitude larger than any modes appearing at other momenta.  In Fig. \ref{fig:Densities_stable} we plot several examples of the multimode non-linear mean field solution, with three distinct peaks clearly visible for $F_p=0.025,\ 0.034, \ 0.050$ [Fig. \ref{fig:Densities_stable} b), c), e)]. In agreement with the linear stability analysis in the previous section, these pump strengths support a stable three-mode ansatz solution at the mean field level, as can be seen in Fig. \ref{fig:OPO_kskykx}.  In contrast, at very low and intermediate pump strengths, the steady-state solutions for the field cannot be described by a three-mode configuration due to the appearance of a high number of significantly occupied states beyond the three-mode ansatz [see Fig. \ref{fig:Densities_stable} a), d)]. These regions in pump strength coincide with the unstable three-mode ansatz shown in Fig. \ref{fig:OPO_kskykx}.

\subsection{Beyond the mean field approximation \label{sec:2Dstochastic}}

Due to the pumping scheme acting in the $x$ direction, the steady-state field configurations presented in the previous section show zero density for modes with $k_y\neq0$, since there is no mechanism to excite these modes within the mean field approximation.  An additional numerical study including fluctuations is therefore required to capture all the details of the system, particularly its behaviour in both the $k_x$ and $k_y$ directions, which is the topic of the present section. We present results for the full multimode non-linear problem with quantum fluctuations described by the stochastic equation given in \mbox{Eq \eqref{eq:sCGPE_pola},} which provides a more complete description of the system than the mean field analysis.  We note that the full stochastic form of Eq \eqref{eq:sCGPE_pola} can also be derived in the semiclassical limit of the Keldysh field theory, from which it can be seen that this method includes effects to all orders in classical fields and up to second order in quantum fluctuations,\cite{PhysRevB.89.134310} making it almost exact for systems with large numbers of particles.  

\begin{figure}[h!]
\includegraphics[width=\columnwidth]{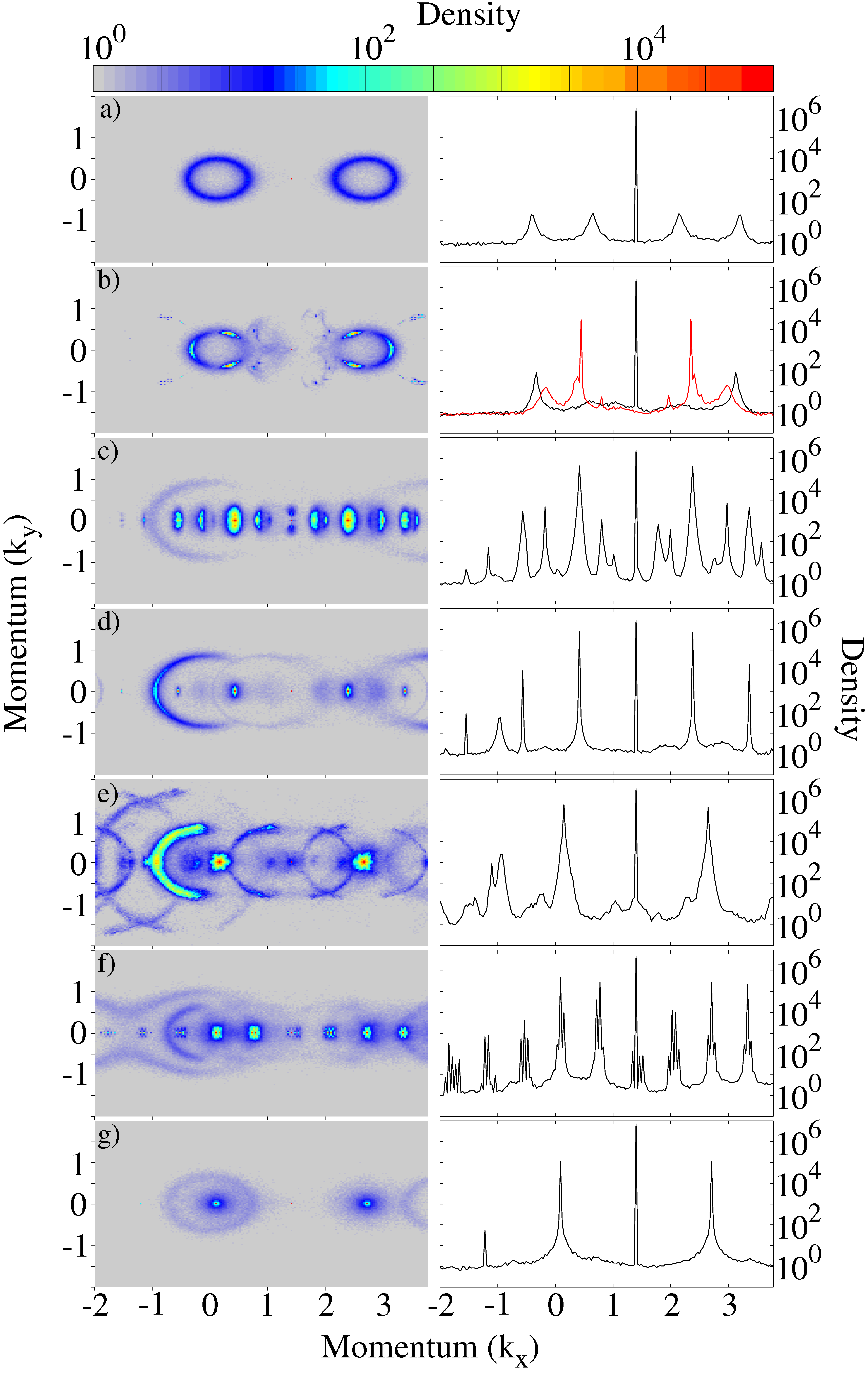}
\caption{ Density of the polariton field $|\psi|^2$ (log scale) at pump strengths across the OPO region. In the left column, we plot $|\psi|^2$ in 2D momentum space, while the profiles through $k_y = 0$ are plotted in the right column. The panels are marked a)-g) by their pump strength $F_p$: a) $F_p = 0.0135$; b) $F_p = 0.016$; c) $F_p = 0.0235$; d) $F_p = 0.026$;  e) $F_p = 0.0335$; f) $F_p = 0.0435$; g) $F_p = 0.051$. For $F_p = 0.016$ the most prominent peaks on the signal and idler rings occur at $k_y=\pm0.327$, so an additional line (red) is plotted in the right hand panel b) showing the density along $k_y=0.327$. The pump state appears in all figures as a peak with maximum density at $\vect{k}=(1.4,0)$.  
\label{fig:TWA_cuts}} 
\end{figure}

\begin{figure}
\includegraphics[width=\columnwidth]{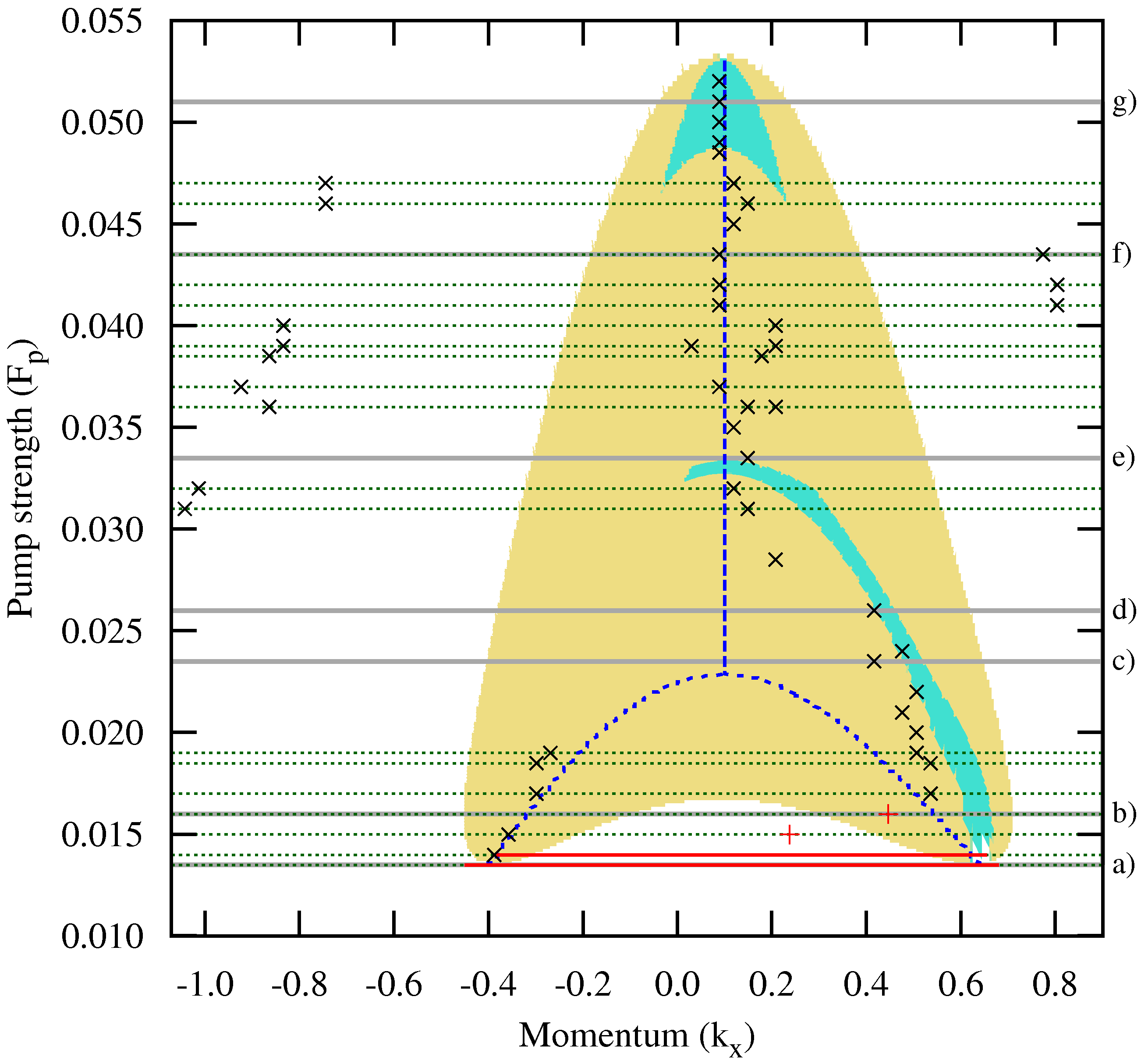}
\caption{ $k_s$ from the multimode stochastic simulations plotted against the stable regions of Fig. \ref{fig:OPO_kskykx}.  Black crosses ($\times$) indicate the locations of the most prominent peaks within this momentum range, which occur along $k_y=0$. Red plusses ($+$) mark where the most prominent peaks have $k_y\neq0$.  Red horizontal lines show the diameter of the rings seen at low $F_p$.  Green dotted lines mark $F_p$ for which a single signal state is not well defined, with multiple crosses at the same $F_p$ indicating several peaks of the same order of magnitude.  Solid grey lines labelled a)-g) mark the profiles plotted in Fig. \ref{fig:TWA_cuts}.  
\label{fig:TWA_v_Stabil}} 
\end{figure}

In Fig. \ref{fig:TWA_cuts} we plot the density of the polariton field in 2D momentum space for different values of external drive, $F_p$, within the OPO region. Fig. \ref{fig:TWA_v_Stabil} gives a comparison between these results and the linear response analysis shown previously. The colour map for the left column of Fig. \ref{fig:TWA_cuts} has been truncated to between $10^{0}$ and $10^5$ to increase the visibility of the main features. As shown in the right column, the main peaks actually have densities above $10^5$ for most values of $F_p$.  

We highlight that the steady-state polariton field, considering quantum fluctuations, shows four different regimes depending on the strength of the external pump, which we now describe in order of decreasing pump strength from the upper threshold.  First, we observe that at high pump strengths, i.e. close to the upper threshold, the polariton field has a density distribution in momentum space with three different significantly occupied modes, the signal, pump and idler (see Fig. \ref{fig:TWA_cuts} g as an example).  This regime coincides in pump strength with the region of stable three-mode mean field ansatz near the upper threshold (turquoise region at high $F_p$ in Fig. \ref{fig:TWA_v_Stabil}).  Additionally, as seen in Fig. \ref{fig:TWA_v_Stabil}, for these high pump strengths the momentum of the signal mode agrees quite well with that found by considering where the single mode description is most unstable (i.e. $k_s = k_s^m$).  

On decreasing the pump strength, we enter a range of pump strengths where the three-mode description of the polariton field fails in almost all cases, as indicated by the green dotted lines in Fig. \ref{fig:TWA_v_Stabil}.  This is due to the appearance of multiple peaks of the same order of magnitude in density, for both the signal ($k_x<k_p$) and idler ($k_x>k_p$) fields, meaning that a single specific value of $k_s$ cannot be defined here.  Additional modes beyond the three-mode description, which are often referred to as satellites,\cite{PhysRevB.71.115301, FMMMHSVorticesinOPOChapt} can become populated by further scattering channels, where signal or idler polaritons scatter from each other or from those in the pump mode.\cite{PhysRevB.64.075311, PhysRevB.65.081308, FMMMHSVorticesinOPOChapt, PhysRevLett.101.136401}  The three-mode description fails here since some of the satellites become comparable in size to the expected signal and idler modes.  Notably, the range of pump strengths at which we observe such solutions corresponds to the range for which the three-mode mean field ansatz has no stable solutions.  While small satellite states are commonly seen in both experimental\cite{PhysRevB.65.081308,PhysRevB.64.075311} and numerical\cite{PhysRevB.71.115301, FMMMHSVorticesinOPOChapt, Gavrilov2007, PhysRevLett.101.136401} studies, the situation where the occupation of these extra modes spontaneously becomes large enough to invalidate the three-mode description, has not yet been observed experimentally. In a few cases, such as that seen in \mbox{Fig. \ref{fig:TWA_cuts} f)} and 
\mbox{Fig. \ref{fig:Densities_stable} f)}, we see a particularly unusual configuration of extra modes, where a peak appears between the pump and the expected signal, as opposed to the usual arrangement, where satellites are evenly spaced by $k_p - k_s$.   

At intermediate to low values of external pump, we observe that the signal and idler fields again develop single sharp maxima at specific values of momentum, allowing a description in terms of three modes.  Examples of this are shown in panels c), d) and e) of Fig. \ref{fig:TWA_cuts}.  In these cases satellite states are still visible on the logarithmic scale, but are at least an order of magnitude smaller than the signal and idler.  Although the chosen $k_s$ is not always within the region where the three-mode mean field ansatz is stable, it seems to adhere more closely to this region than to the $k_s^m$ from the single-mode linear response analysis. 

At very low pump strengths ($F_p<0.02$), the three-mode description once again fails, and we instead see the signal and idler fields having multiple maxima distributed around a ring in momentum space. At the higher of these $F_p$, the main maxima still occur along $k_y = 0$, but at lower pump strengths maxima at $k_y \neq 0$ become the most relevant as shown in Fig. \ref{fig:TWA_cuts} b).  Very close to the lower threshold, Fig. \ref{fig:TWA_cuts} a) $F_p = 0.0135$, the distinct maxima give way to a nearly uniform density for states on the rings.  We note that the size of these rings matches almost exactly with the rings of $k_s^m$ from the linear stability analysis at the same pump strengths (e.g. Fig. \ref{fig:circle}).  Such rings have previously been observed in calculations of the luminescence of the polariton field at the OPO lower threshold,\cite{paper1} and have also been seen in experiments in the bistable regime, although in those cases the rings intersect at the pump momentum to form a figure of eight pattern.\cite{PhysRevB.70.205301,PhysRevB.77.115336}

\section{Exciton-photon model \label{XCmodel}}

In this section we extend our study by considering the exciton-photon model, which describes the polariton system in terms of the constituent fields
~\cite{PSSB:PSSB200560961, PhysRevB.63.041303, PhysRevB.92.035307, PhysRevB.72.125335} (with $\hbar = 1$):
\beqy
i\partial_t \bpm \psi_x \\ \psi_c \epm = \hat{H}_{xc}\bpm \psi_x \\ \psi_c
\epm + \bpm 0 \\ f_c(\mathbf{r},t) \epm 
\label{eq:ex_ph_model}
\eeqy
where
\beqs
\hat{H}_{xc} = \bpm \omega_x(-i\nabla) - i \kappa_x +
g_\mathrm{x}|\psi_x|^2 & \Omega_R/2 \\ \Omega_R/2 &
\omega_c(-i\nabla) - i \kappa_c \epm.
\eeqs
Here $\psi_x$ and $\psi_c$ are the exciton and photon fields respectively, we assume a constant exciton-exciton interaction $g_{\mathrm{x}}$,~\cite{PSSB:PSSB200560961, RevModPhys.82.1489, SemicondSciTech.18.279} and $\omega_x, \omega_c$ correspond to the dispersion relations of the exciton and photon fields respectively.~\cite{RevModPhys.82.1489, Microcavities} Since the exciton mass is much larger than the cavity photon mass, its dispersion is considered to be flat.~\cite{FMMMHSVorticesinOPOChapt, RevModPhys.82.1489,keeling} The exciton and photon decay rates are given by $\kappa_{x}$, $\kappa_c$ respectively and are considered to be equal in all calculations, with values $\kappa_x = \kappa_c = 1/\tau_c \approx 0.045$, where $\tau_c \approx 5.8\ \mathrm{ps}$. We use $\Omega_R = 4.4\ \mathrm{meV}$ and $g_X = 0.002\ \mathrm{meV \mu m ^2}$.  The external pump $f_c$ is again a plane wave of the form $F'_p e^{-i\omega_pt}\delta_{\mathbf{k},\mathbf{k}_p}$. The numerical integration runs to $72\ \mathrm{ns}$ in all cases. 

First we consider the steady-state mean field solution for a single mode (plane wave ansatz) for both the exciton and the photon fields, which coincides with the pump energy and momentum, 
i.e. $\psi_c(\vect{r})^{} = P_c e^{-i\omega_pt - i\vect{k}_p\vect{r}}, \psi_x(\vect{r})^{} = P_x e^{-i\omega_pt - i\vect{k}_p\vect{r}}.$  We substitute these expressions into the Eq. \eqref{eq:ex_ph_model}, 
and, after rearranging, obtain $P_c = \frac{2}{\Omega_R}(\omega_p +i\kappa_x-\omega_x -g_{\mathrm{x}}|P_x|^2)P_x$ from the first line.  So, although the external pump is written in terms of both 
$P_c$ and $P_x$, $P_c$ can be eliminated and the exciton occupation found directly from $F_p' = (\omega_p +i\kappa_c -\omega_c(\mathbf{k}_p))P_c-\frac{\Omega_R}{2}P_x$.~\cite{PSSB:PSSB200560961}  
Adding fluctuations to both the exciton and photon fields, leads to the linear response matrix:~\cite{PSSB:PSSB200560961, PhysRevB.72.125335}
\beq
L_{xc} = \bpm A(\Delta \mathbf{k}) & B \\ -B^* & -A^*(-\Delta \mathbf{k}) \epm \label{XCLRmatrix} 
\eeq  
with 
\beqs
A(\Delta \mathbf{k}) = \bpm \omega_x+ 2g_{\mathrm{x}}|P_x|^2 & \underline{\Omega_R} \\
  - \omega_p -i\kappa_x  & \! 2 \\
\underline{\Omega_R} & \omega_c(\mathbf{k}_p+\Delta \mathbf{k}) \\
\! 2 &  - \omega_p-i\kappa_c \epm \label{XCLRdiags}
\eeqs 
and 
\beqs
B = \bpm g_{\mathrm{x}}{P_x}^2 & 0 \\ 0 & 0 \epm. \label{XCLRoffdiags} 
\eeqs
The eigenvalues of Eq. \eqref{XCLRmatrix} are calculated numerically and the maximum value of the imaginary part at each pump strength extracted as in section \ref{sec:one_mode_pola}.

\subsection{Stability analysis and mean field numerics}

\begin{figure}[t!]
\includegraphics[width=\columnwidth, trim = 0 6 6 5, clip]{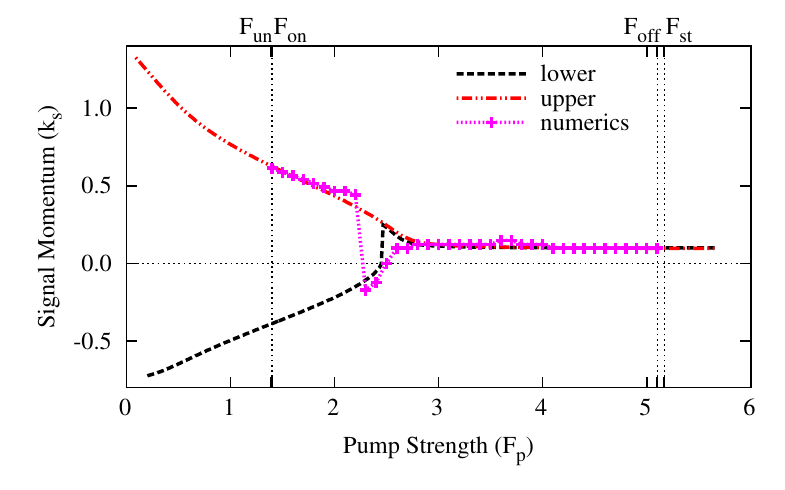} 
\caption{  $k_s$ from numerical integration of the coupled cGPEs (magenta dotted with $+$) and upper and lower values of $k_s^m$ from linear response (red dash-dotted and black dashed respectively).  
\label{fig:XCcomparisons}} 
\end{figure}

First, taking $k_p = 1.4$ (in dimensionless units) and considering the pump tuned resonantly to the lower polariton dispersion, we can see in Fig. \ref{fig:XCcomparisons}  qualitatively similar behaviour to that described using the simplified  lower polariton model over much of the OPO regime. 
\begin{figure}[h!]
\includegraphics[width=\columnwidth]{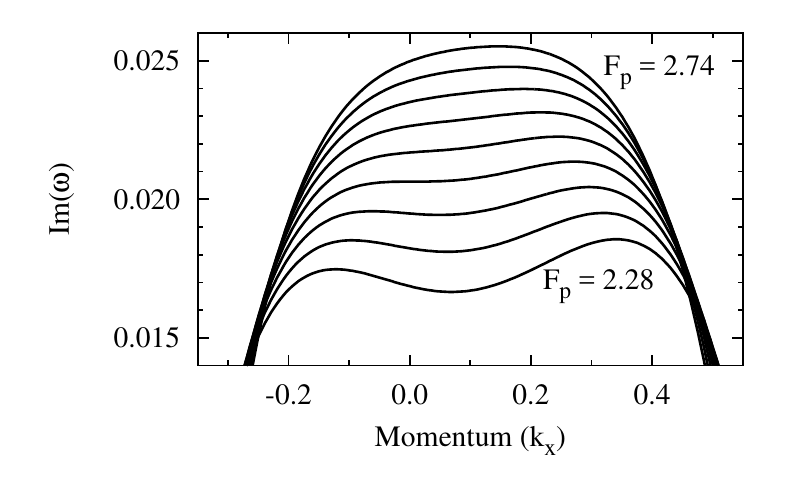}
\caption{Imaginary part of one of the eigenvalues of Eq. \eqref{XCLRmatrix} ($\Im(\omega^+)$) near $k=0$ for a range of pump strengths near the sharp jump in $k_s$ that occurs when the two peaks combine.\label{fig:XCimsdetail}} 
\end{figure}
At higher pump strengths, there is a single $k_s$ value, $k_s\approx 0.1$, over a significant range of pump strengths.  At lower pump strengths the system exhibits two distinct values of $k_s$ in the $k_y = 0$ plane, appearing both in the numerics and the stability analysis. In contrast to the lower polariton model, the approach to the coalescence point is abrupt. This is the result of the dip between the two peaks disappearing as shown in Fig. \ref{fig:XCimsdetail}.

\subsection{Changing the pump properties.\label{extensionsvarkp}}

\begin{figure}[t!]
\includegraphics[width=\columnwidth]{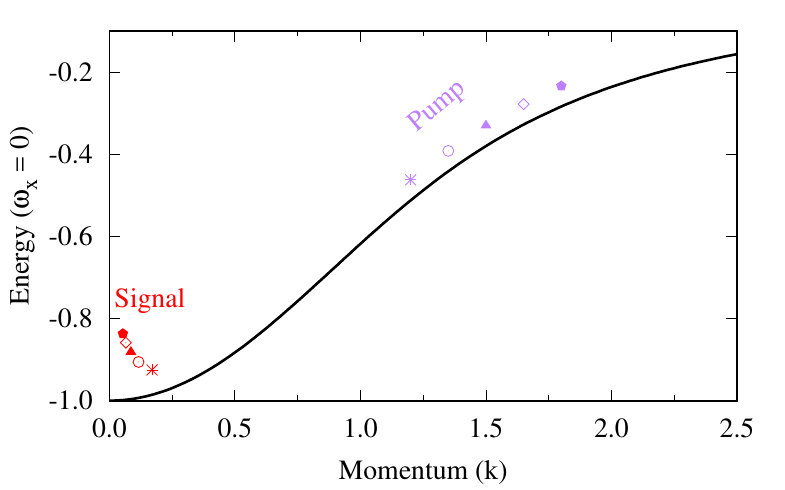} 
\caption{ Signal energy [$\Re(\omega^+)$] and momentum $k_s$ with pump energy $\omega_p$ and momentum $k_p$ (corresponding symbols) for $\Delta = +0.05$.  The signal used is from the `coalescence' point.  
The solid line is the unperturbed lower polariton dispersion. \label{fig:signalenergy}} 
\end{figure}

In this section we study the behaviour of $k_s^m$ obtained from the stability analysis when changing the detuning $\Delta\equiv\omega_p -\omega_{lp}(k_p)$ and the momentum $k_p$ of the external pump.  
\begin{figure}[b!]
\includegraphics[width=\columnwidth]{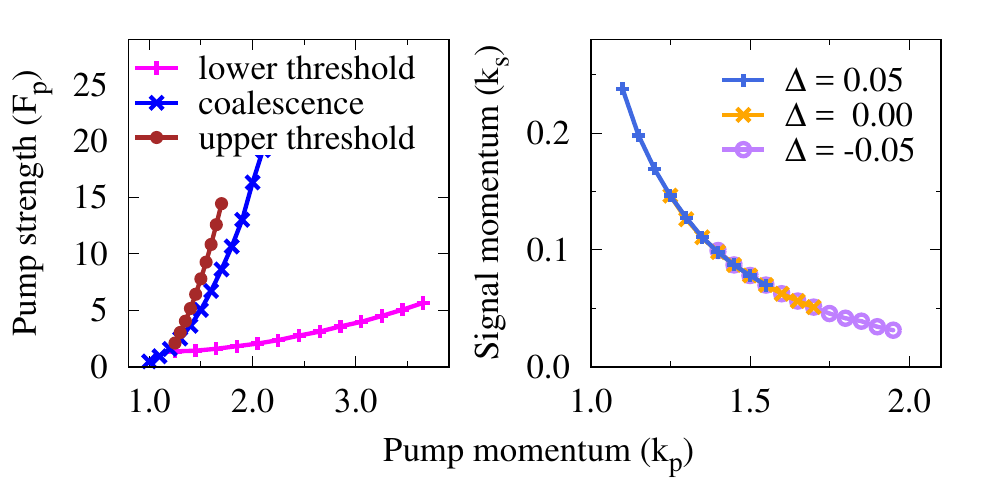}
\caption{ Left: Pump strengths of the upper (brown dots) and lower (magenta $+$) thresholds and coalescence point (blue $\times$) for varying $k_p$, with $\omega_p = \omega_{lp}(k_p)$ ($\Delta = 0$).  
Right: signal momentum at the upper threshold with varying $k_p$ for three detunings, $\Delta = -0.05, 0.00, +0.05$.\label{fig:kswithkp}} 
\end{figure} 
Firstly, in the left hand side of Fig. \ref{fig:kswithkp}, we plot the various thresholds that we have been interested in for a range of pump momenta. We see that all thresholds increase with $k_p$.~\cite{KDunnettThesis}  In the right hand panel of Fig. \ref{fig:kswithkp}, we show that changing the detuning of the pump in either direction, while remaining within the optical limiter regime, has no noticeable effect on the value of $k_s$ for a given $k_p$.  The signal momentum decreases with increasing $k_p$ (see Fig. \ref{fig:signalenergy}), however, the change in $k_s$ is small compared to the corresponding change in $k_p$.  The independence of $k_s$ on the detuning is consistent with experiments, while the variation in signal momentum is probably too small to be observed in current experiments.~\cite{PhysRevB.68.115325, EPL.67.997}  We also investigate the behaviour of the \emph{signal energy} (the real part of the eigenvalue of the linear response matrix at $k_s$) as a function of the external pump momentum, $k_p$, at $\Delta = +0.05$ (see Fig. \ref{fig:signalenergy}). We observe that the signal energy is monotonically increasing with $k_p$, consistent with experiments.~\cite{PhysRevB.68.115325}

\section{Conclusions \label{Conclusions}}

In this work we have investigated the rich phenomenology of the signal state in the parametrically driven polariton system.  In particular, we focused on determining in which regime the commonly used three mode OPO description is valid, and whether we can predict the signal momentum from a simple stability analysis, without the need to resort to the full solution of non-linear multi-mode equations. In all cases, we restricted our analysis to the optical limiter regime, where the pump mode occupation is a monotonic function of the external pump strength.  

Our non-linear stochastic simulations of the polariton OPO show behaviours at different pump strengths that can be classified into four cases: i) three-mode solutions near the upper threshold, ii) multi-mode solutions (large satellites in addition to the signal, pump and idler modes) at higher intermediate pump strengths, iii) approximately three-mode solutions at lower intermediate pump strengths (small satellites), and iv) rings near the lower threshold.  The three-mode ansatz is a good approximation in cases i) and iii) but cannot represent the form of the OPO seen in cases ii) and iv) well.  The range of pump strengths for case ii) corresponds closely to where the three-mode mean field ansatz has no stable solutions. 

Our study of the momentum of the most unstable eigenvalues from the linear response analysis of the pump-only case ($k_s^m$), compared with the $k_s$ chosen by the system described by the full multi-mode non-linear stochastic equations, shows that linear response analysis provides a good estimate for both the $k_s$ value chosen near the upper threshold [case i)] and the size of the rings that occur near the lower threshold [case iv)], but is less useful for intermediate pump strengths.  We extended our study using the more complete exciton-photon model, and find that similar conclusions hold. 

Finally, we have shown that the signal momentum does not depend on the detuning of the pump laser and varies little with the pump momentum, while the signal energy does increase with the pump energy and momentum in agreement with previous experiments.~\cite{PhysRevB.68.115325}  

\acknowledgments{ We thank Th. K. Mavrogordatos, F. M. Marchetti and A. Berceanu for helpful discussions. K. D. acknowledges help from J. M. Fellows with early versions of the Fortran code used. We acknowledge support from EPSRC (grants EP/I028900/2 and EP/K003623/2).}

\end{document}